\documentclass[prb,preprint,showpacs,preprintnumbers,amsmath,amssymb,showkeys]{revtex4}
\usepackage{epsfig}
\usepackage{multirow}
\usepackage{rotating}

\begin{document}

\title{Co dimers on hexagonal carbon rings proposed as subnanometer magnetic storage bits}

\author{Ruijuan Xiao}

\author{Daniel Fritsch}

\author{Michael D. Kuz'min}

\author{Klaus Koepernik}

\author{Helmut Eschrig}

\author{Manuel Richter}

\affiliation{IFW Dresden e.V., PO Box 270116, D-01171 Dresden, Germany}

\author{Knut Vietze}

\author{Gotthard Seifert}

\affiliation{Physikalische Chemie, Technische Universit\"{a}t Dresden, D-01062 Dresden,
Germany}

\date{\today}

\begin{abstract}
It is demonstrated by means of density functional and {\it ab-initio}
quantum chemical calculations,
that transition metal - carbon systems have 
the potential to enhance the presently achievable
area density of magnetic recording 
by three orders of magnitude.
As a model system, Co$_2$-benzene with a 
diameter of 0.5 nm is investigated.
It shows a magnetic anisotropy in the order of 0.1 eV per molecule,
large enough to 
store permanently one bit of information at temperatures
considerably larger than 4 K.
A similar performance can 
be expected, if cobalt dimers are deposited on graphene or on graphite.
It is suggested that the subnanometer bits can be written by simultaneous application of
a moderate magnetic and a strong electric field.
\end{abstract}

\pacs{31.15.es, 75.30.Gw, 75.75.+a}

\keywords{applications of density functional theory, magnetic anisotropy, magnetic properties of nanostructures}

\maketitle

Long-term magnetic data storage 
requires that spontaneous magnetization reversals should occur 
significantly less often than once in ten years.
This implies that the 
total magnetic anisotropy energy (MAE)
of each magnetic particle should exceed
40 $kT$,~\cite{Charap_IEEETransMagn33_978}
where $k$ is the Boltzmann constant and $T$ is the temperature. 
Among the elemental ferromagnets (Fe, Co, Ni, and Gd), cobalt metal 
shows the highest MAE, about 0.06 meV per atom in the hexagonal
close packed structure.
Thus, Co is the main ingredient of magnetic data storage materials at present.
At room temperature,
data loss due to fluctuations is avoided, if
a Co grain contains not less than 
40 $k\cdot300$ K$/0.06$ meV $\approx$ 15,000 atoms.
In fact, the grain diameter of contemporary Co(Cr,Pt,SiO$_2$)
recording media is close to 8 nm, each grain containing about 
50,000 atoms and each bit being composed of some dozen grains.~\cite{Thomson}
The grain size could be considerably reduced by using the intermetallic compounds 
FePt or CoPt with record MAE of almost 1 meV per atom in their structurally 
ordered L1$_0$ bulk phase. It is, however, hard to achieve the 
required ordered structure in nano-particles.~\cite{Gruner_prl100_087203}

Obviously, a further reduction of the bit size is primarily limited by the 
value of MAE per atom. Recent efforts to enhance this value were focused 
on single atoms or small clusters deposited on the surface of heavy metals. 
This approach combines two ideas:~\cite{Bruno_prb39_865}
Firstly, the magnitude of MAE is related to the size of the orbital moments. 
The latter are quenched for highly coordinated atoms but can be large if the 
coordination is low. Secondly, the magnitude of MAE is related to the strength 
of spin-orbit coupling which grows with atomic number. Considerable progress 
was achieved in this way by deposition of single Co atoms on a Pt surface, yielding a 
record MAE of 9 meV per Co atom.~\cite{Gambardella_Science300_1130} Unfortunately, clusters 
of several Co atoms on Pt show a much smaller MAE per atom, roughly 
inversely proportional to the number of 
atoms.~\cite{Gambardella_Science300_1130}

More recently, the magnetic properties of transition metal dimers 
came into the focus of 
interest.~\cite{Strandberg_NatureMaterials6_648,Strandberg_prb77_174416,Fernandez-Seivane_prl99_183401,Fritsch_JCompChem29_2210}
Isolated magnetic dimers are the smallest chemical objects that possess 
a magnetic anisotropy as their energy depends on the relative orientation 
between dimer axis and magnetic moment. Huge MAE values of up to 100 meV 
per atom were predicted by density functional (DFT) calculations for the cobalt 
dimer.~\cite{Strandberg_NatureMaterials6_648,Strandberg_prb77_174416,Fritsch_JCompChem29_2210}

Transition metal dimers can routinely be produced by sputtering and subsequent 
selection using mass-spectroscopy.~\cite{Lombardi_ChemRev102_2431} 
It is 
impossible, however, to access the huge MAE of dimers 
unless they are  bound to some medium. Chemical bonding, on the other hand, 
frequently spoils magnetism. A hitherto unanswered question is,
if one can find a substrate that does not
deteriorate the dimer MAE. In the following, we will demonstrate 
by DFT and by {\em ab-initio} quantum chemical calculations
that carbon is a  suitable host for subnanometer magnetic storage bits
consisting of transition metal dimers with exceptional magnetic anisotropy
in the order of 0.1 eV,
and that such bits can be written by simultaneous application
of a moderate magnetic and a strong electric field. 

An all-electron full-potential local-orbital scheme, 
FPLO-8.00-31,\cite{Koepernik_prb59_1743,FPLO} was employed for the 
DFT calculations.
The valence basis set 
comprised 3d-transition metal (3s, 3p, 3d, 4s, 4p, 4d, 5s), 
4d-transition metal (4s, 4p, 4d, 5s, 5p, 5d, 6s), carbon 
(1s, 2s, 2p, 3s, 3p, 3d), and hydrogen (1s, 2s, 2p) states.
The presented data were obtained using the generalized gradient
approximation (GGA) with a parameterized 
exchange-correlation functional according to Ref. \onlinecite{Perdew_prl77_3865}
A temperature broadening parameter of 100 K for the level 
occupation was used in the cluster mode and
a mesh of $30\times 30\times 1$ {\bf k}-points in the full Brillouin zone in the periodic mode (supercells with 26 \AA ~height).
Geometry optimization was carried out with a scalar relativistic scheme. 
The magnetic anisotropy energy was obtained by means of self-consistent 
full-relativistic calculations. Resulting GGA data were regarded 
as a lower bound for the expected MAE value. To estimate an upper bound, 
an orbital polarization correction (OPC)
\cite{Eriksson_prb41_7311,Eschrig_EurophysLett72_611} in the 
version proposed in Ref. \onlinecite{Eriksson_prb41_7311}
was applied to the cobalt 3d-orbitals.

Co dimers have singly occupied two-fold degenerate 3d-$\delta$
orbitals which are responsible
for the large dimer MAE.\cite{Strandberg_NatureMaterials6_648,Strandberg_prb77_174416}
This degeneracy is not lifted in a hexagonal environment.
Thus, we have chosen to
investigate the interaction between Co$_2$ and benzene (Bz, C$_6$H$_6$).
Previous calculations demonstrated 
that the interaction of transition metals with Bz suitably models their 
adsorption onto graphite.\cite{BelBruno_SurfSci577_167}

Co$_2$C$_6$H$_6$ has an even number of electrons and hence a spin
magnetic moment of an even number of Bohr magnetons.
To find out the lowest-energy geometry and spin magnetic state, five 
possible high-symmetry structures (Fig. 1) were optimized for each of the 
possible values of total spin moment, $\mu_s$ = 0, 2, 4, and 6 $\mu_{\rm B}$.
Different initial spin arrangements (ferro- and ferri-magnetic) 
were considered. The point group symmetry (C$_{6v}$, C$_{2v}$, or 
C$_{\infty}$) as well as the geometry of the Bz were fixed.
We find a binding energy of 1.39 eV for the Co$_2$ dimer to 
the Bz and a perpendicular orientation of  the dimer axis to the 
Bz plane in the ground-state structure. 
We are not aware of previous investigations for
this structure of Co$_2$C$_6$H$_6$.
Perpendicular arrangement of transition metal dimers 
on Bz was evidenced before by spectroscopy on Pd$_2$C$_6$H$_6$ and 
Pt$_2$C$_6$H$_6$.\cite{Luettgens_JChemPhys114_8414} Also, 
gold dimers were predicted to prefer a perpendicular orientation 
on graphene.\cite{Varns_jpcm20_225005}

DFT calculations are known to yield, in most cases, 
binding energies somewhat larger than related experimental values. Thus, 
to confirm the qualitative validity of the energies, structure sequence,
and spin states
of Fig. 1, we performed {\em ab initio} quantum chemical calculations. 
Binding energies were obtained from second order M\o{}ller-Plesset (MP2)
perturbation theory as implemented in Gaussian03.\cite{gaussian}
The Co atoms were described by a scalar-relativistic effective
core potential (ECP) replacing 10 core electrons (MDF10),\cite{Dolg87}
with the corresponding (8s7p6d1f)/[6s5p3d] GTO basis set.
For Bz the standard Dunning correlation-consistent
double-zeta basis set (cc-pVDZ)\cite{Dunning89} was used.
All MP2 energies were evaluated from single point calculations
using the DFT-derived geometries, except for the Co dimer.
For the latter, the MP2 interatomic distance
turned out slightly shorter (0.1909 nm) than the DFT
result (0.1997 nm).
Nonetheless, the binding energy of the Co dimer to Bz
was found to be yet higher (2.67 eV) than in the DFT calculation (1.39 eV).
The perpendicular orientation of the dimer axis to the Bz plane
was found to yield the lowest energy.
The other configurations are separated from this structure
by 1.40 eV, 2.11 eV, and 2.28 eV, respectively, in the sequence depicted
in Fig. 1.
The same ground state spin magnetic moments as in the DFT calculations
were found with the exception of the uppermost configuration,
were a total moment of 2 $\mu_{\rm B}$ was obtained by MP2
compared to 4 $\mu_{\rm B}$ obtained by DFT.
Thus, the quantum chemical calculations confirm the main DFT result,
that bonding of a Co dimer with a single molecule of Bz results
in the structure depicted at the bottom of Fig. 1 with a total spin
magnetic moment of 4 $\mu_{\rm B}$.

A compilation of related single electron level schemes is given in Fig. 2. 
The rightmost panel shows the textbook electronic structure of Bz, 
while the leftmost panel refers to the free Co$_2$, as recently discussed in 
Refs. \onlinecite{Strandberg_NatureMaterials6_648,Strandberg_prb77_174416}. 
The two inner panels show our new results for the ground-state structure 
of Co$_2$Bz and for CoBz, included for comparison. 

The most important feature of the Co$_2$ level scheme is a 
two-fold degenerate singly occupied 3d-$\delta_{u}^{*}$ state. It is split 
by spin-orbit interaction (not shown),
if the magnetic moment is oriented along the 
dimer axis but stays degenerate if the moment is perpendicular to the 
axis.\cite{Strandberg_NatureMaterials6_648,Strandberg_prb77_174416}
The related energy difference, the MAE, was predicted to be 50 to 60 meV 
per molecule.\cite{Strandberg_NatureMaterials6_648,Strandberg_prb77_174416,Fritsch_JCompChem29_2210}
This value should be considered as a lower estimate (L) to the expected 
MAE since it was evaluated without allowing for the so-called orbital 
polarization corrections.
Rather an upper estimate (U), 188 meV per
molecule,\cite{Fritsch_JCompChem29_2210}
is obtained by including OPC.

Surprisingly, it turns out that bonding of Co$_2$ on Bz does not 
lead to any deterioration of its magnetic properties, but 
rather improves them: The spin moment stays 4 $\mu_{\rm B}$, 
as in the free dimer. Noteworthy, any other spin state, including an 
anti-ferromagnetic solution with about 2 $\mu_{\rm B}$ on each Co atom, 
has a much higher energy, at least 800 meV above the ground state. 
As a consequence, the 3d-$\delta_{u}^{*}$ level is still singly occupied, 
Fig. 2. It is now further separated from the other levels than in the 
free dimer. Thus, while the lower 
estimate to MAE is hardly changed (51 meV per molecule),
the upper one (330 meV per molecule) is even higher than in the free dimer.
This huge MAE is accompanied by a large ground 
state orbital moment, about 2 $\mu_{\rm B}$ per molecule. 
The latter is largely suppressed if the total magnetic moment 
is directed parallel to the Bz plane. For this orientation, the 
3d-$\delta_{u}^{*}$ level is split by a tiny Jahn-Teller distortion 
of the Bz with hardly any influence on the MAE.

The CoBz molecule was earlier investigated both 
experimentally~\cite{Kurikawa_Organometallics18_1430} and by quantum chemical 
calculations,\cite{Bauschlicher_JPhysChem96_3273} but its magnetic behavior 
deserves further studies. Fig. 2 shows the CoBz level scheme for its 
lowest spin state in the C$_{6v}$ geometry with the Co atom 0.1501 nm 
above the center of the Bz ring. Here a 3d-$\pi$ doublet is singly occupied. 
Spin-orbit splitting of 3d-$\pi$ levels is smaller than that of 3d-$\delta$ 
levels. Hence, we find somewhat smaller MAE values of 20 (L) and 90 (U) meV 
per Co atom. These numbers are comparable with the recently predicted MAE of 
7.5 meV per tantalum atom (L) for TaBz.\cite{Mokrousov_Nanotechnology18_495402}

Technological application of Co dimers might 
proceed not via the deposition of Co$_2$Bz on a substrate, but rather via 
deposition of Co$_2$ on purified graphite. With this end in view, we have 
evaluated the bonding and magnetic properties of regular 3$\times$3 arrangements 
of Co and of Co$_2$ on graphene as a 
model for the graphite (0001) surface. The results resemble those 
described for Bz:
(i) the perpendicular orientation of Co$_2$ above the center of a carbon 
hexagon is preferred against any parallel orientation, including a position of 
the two Co atoms above two adjacent hexagons; 
(ii) single Co atoms prefer the hollow position as well; 
(iii) the binding energy is about 15\% smaller than in the Bz case; 
(iv) both single Co atoms and Co dimers keep their essential electronic 
and magnetic features. Co dimers show a magnetic anisotropy 
(L: 48, U: 320 meV per dimer) very similar to the Bz case, while the 
MAE of single Co atoms on graphene is somewhat reduced (L: 5, U: 40 meV 
per atom) in comparison to CoBz. 

Fig. 3 demonstrates the strong sensitivity of Co MAE to its environment and 
compares L and U to MAE with experimental data. While the U to MAE of 
Co$_2$ on C (0001), once confirmed by experiment, could guarantee 10-year 
stability of an information bit at the temperature of liquid nitrogen, 
$T \approx 80$ K, even the lower estimate is a factor of five higher 
than the measured MAE of a Co dimer on a Pt 
surface~\cite{Gambardella_Science300_1130} and would provide the required 
stability at 12 K.
A recent experiment
demonstrated that single-molecule magnets can
indeed be used for magnetic data storage: for Fe$_4$-organometallic
complexes with an MAE of about 2 meV per molecule,
an information decay time of about 200 s was measured
at a very low temperature, 0.5 K.\cite{Mannini09}

Among all T$_2$-Bz molecules investigated in the present work
(T = Fe, Co, Ni, Ru, Rh, Pd),
Ru$_2$Bz is the only other interesting candidate for high MAE 
application. Ru$_2$
shows similarly good bonding behavior with Bz (energy and 
geometry) as the cobalt dimer and Ru$_2$Bz has 
an even higher L to MAE of 100 meV 
per molecule. However, the related magnetic ground state has a lower moment 
(2 $\mu_{\rm B}$) and lies only 250 meV below a state with zero moment. 
This situation is similar to a recently predicted effect termed 
{\textquotedblleft colossal magnetic anisotropy\textquotedblright} in 
monatomic platinum nanowires~\cite{Smogunov_NatureNanotechnology3_22} 
and deserves further investigation.

The preparation of transition metal dimers on C(0001) and the
experimental verification 
of their extraordinary properties, e.g., by single-atom spin-flip 
spectroscopy,\cite{Heinrich_Science306_466} will be an interesting 
goal on its own. Technological use would require to solve at least three 
additional problems: fabrication of large regular arrays;
protection against oxidation without reducing the anisotropy; 
new read/write technologies. 
Let us finally discuss a possible method to solve the latter problem.
Conventional write technology makes use of magnetic fields B in the
order of 1 T, Ref.~\onlinecite{Thomson}. 
It would fail in the present
situation, where a field B = MAE/$\mu_s$
of several hundred tesla would be needed.
We thus propose to combine a moderate quasi-static magnetic field
with a strong electric field pulse applied, e.g., through the tip
of a scanning tunneling microscope (STM), Fig. 4.
Before the electric pulse is applied, Fig. 4a, the magnetic field
creates a metastable situation but the direction of magnetization
is protected by the MAE barrier.
An electric field of about $5\cdot 10^{9}$ V/m, Fig. 4b,
produces a crossing of the half occupied 3d-$\delta_{u}^{*}$ level with
the lowest empty carbon-related level (denoted $\pi_4^*,\;\pi_5^*$
in the second panel of Fig. 2). As a result, the upper Co atom,
hosting the main weight of the 3d-$\delta_{u}^{*}$ orbital, is 
ionized on a time scale of $10^{-14}$ s and the electron
is absorbed by the reservoir. For comparison, the time scale 
for electron tunneling from an STM tip is typically $10^{-10}$ s.
Our DFT calculations for Co$_2^+$Bz show that both the structure
and the spin magnetic state (now with an enhanced spin moment of
5 $\mu_{\rm B}$) are robust. The magnetic anisotropy
is however virtually quenched due to the ionization.
Hence, the direction of magnetization can now be flipped
(e.g., triggered by electromagnetic radiation) in
the magnetic field, Fig. 4c.
Eventually, the electric field is switched off, an electron supplied
by the reservoir restores charge neutrality of the Co$_2$Bz, Fig. 4d,
and the MAE barrier is recovered.

Summarizing, we predict that bonding of Co dimers
on hexagonal carbon rings like benzene or graphene results in 
a perpendicular arrangement of the dimers with respect to the
carbon plane and in a magnetic ground state. 
In this structure, a division of tasks takes place: while the
Co atom closer to the carbon ring is responsible for the
chemical bonding, the outer Co atom hosts the larger share
of the magnetic moment.
The huge magnetic anisotropy of the free dimer is preserved
in this structure, since the degeneracy of the highest occupied
3d-$\delta$ orbital is not lifted in a hexagonal symmetry.
Thus, it should be possible to circumvent the hitherto favored use of heavy metal
substrates to achieve large magnetic anisotropies.
On the contrary, robust and easy-to-prepare 
carbon-based substrates are well-suited for this task.
Once confirmed,
the present results may constitute an important step towards
a molecular magnetic storage technology.

We thank Y. Dedkov, U. Nitzsche, and U. R\"o\ss{}ler
for discussion. The Deutsche 
Forschungsgemeinschaft (SPP1145 and FOR520)
is acknowledged for financial support. 

\clearpage

\begin{figure}
\epsfig{file=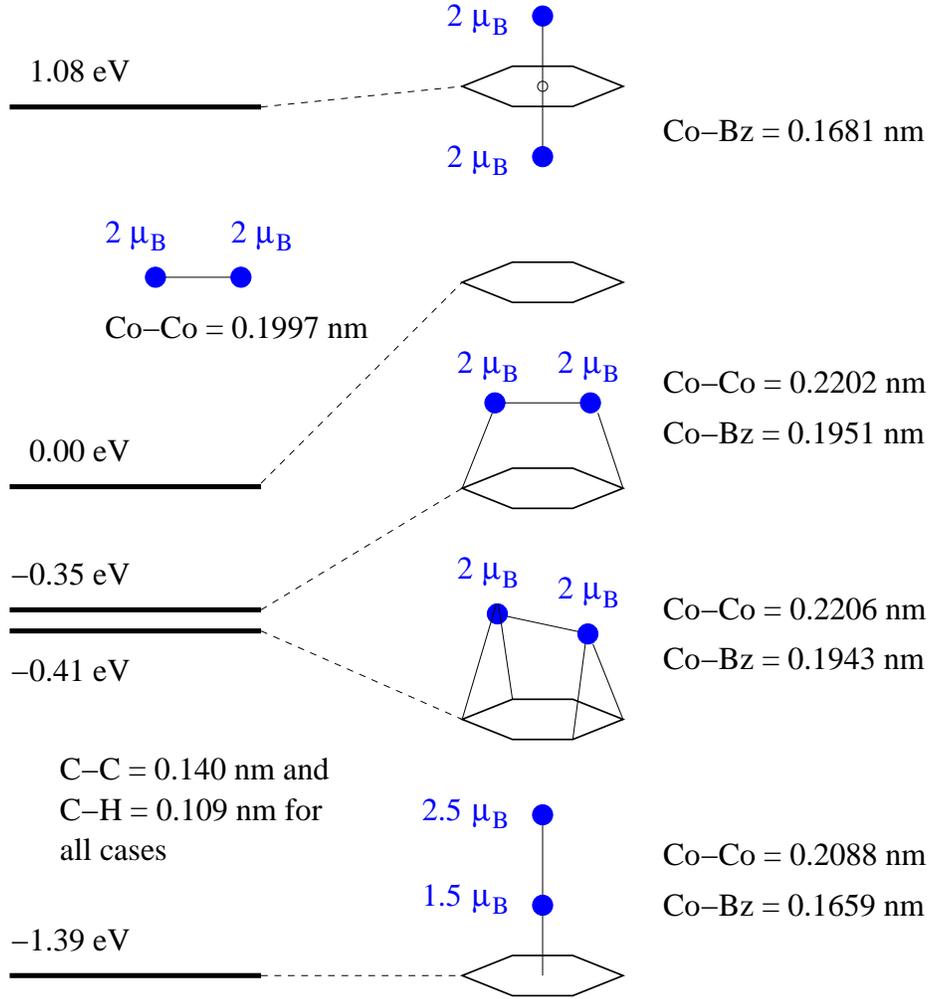, width=0.75\textwidth}
\caption{(color online)
DFT energies, interatomic
distances, and ground-state spin magnetic moments
of different Co$_2$-benzene configurations.
The energies are given relative to the energy sum of free Co$_2$
and free Bz. Hexagons and blue bullets indicate Bz and
Co atoms, respectively. Co-Bz denotes the distance between the
Bz plane and the nearest Co atom.
}
\label{fig1}
\end{figure}

\clearpage

\begin{figure}
\epsfig{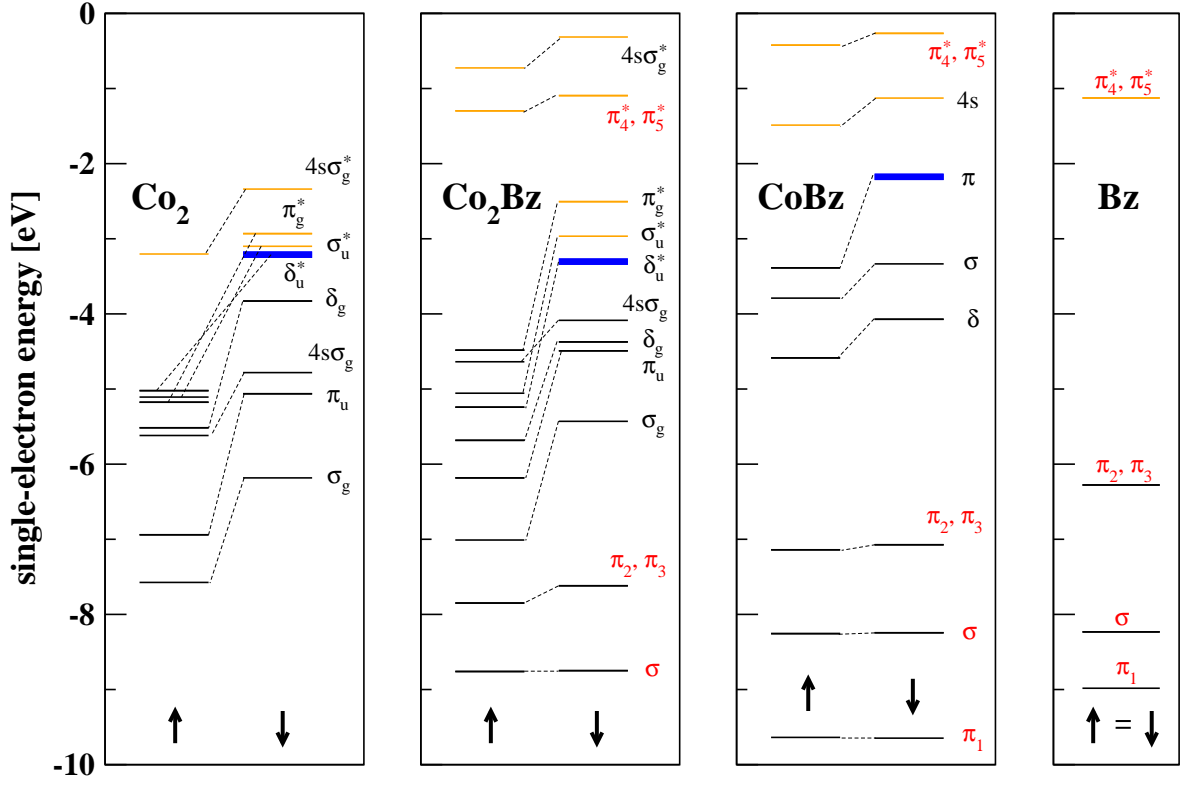}
\caption{(color online)
Scalar-relativistic DFT single-electron
levels of Co$_2$ (left panel), Co$_2$Bz (ground-state structure,
second panel), CoBz (third panel), and Bz (right panel, Bz = C$_6$H$_6$).
All energies refer to a common vacuum level. Black lines
denote occupied states, orange lines denote empty states, and
thick blue lines indicate singly occupied two-fold degenerate
states. With the exception of Bz, the levels are spin-split
($\mu_s$ = 4, 4, and 1 $\mu_{\rm B}$ for Co$_2$, Co$_2$Bz, and CoBz,
respectively). Majority states are indicated by up-arrows, minority
states by down-arrows. Co-like states are labelled in black
and Bz-like states are labelled in red.
}
\label{fig2}
\end{figure}

\clearpage

\begin{figure}
\epsfig{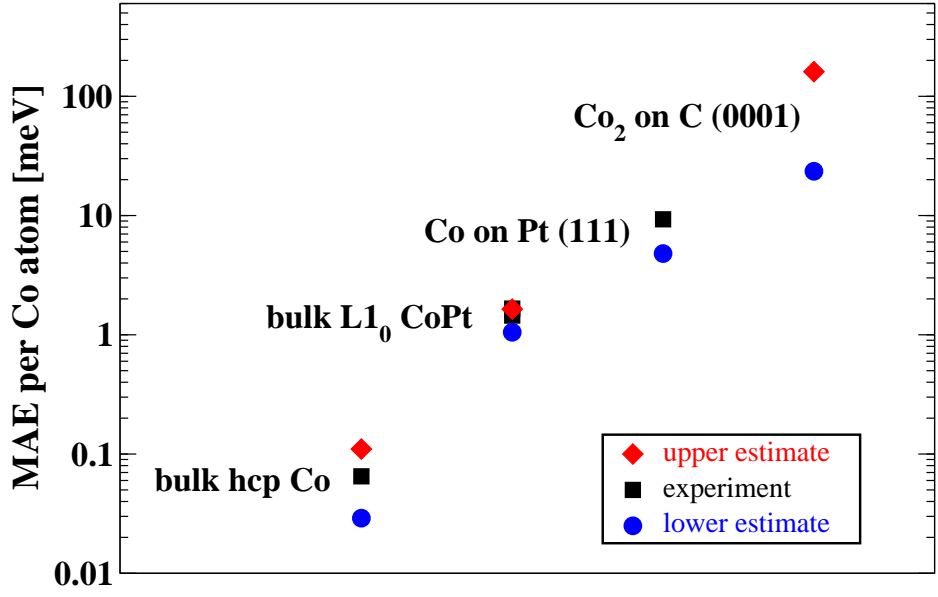}
\caption{(color online)
Magnetic anisotropy energy of Co atoms
in different chemical and structural environments. Black squares
denote experimental data, blue circles and red diamonds denote
lower and upper estimate theoretical data, as introduced in the
text. Bulk hcp Co,\cite{Trygg_prl75_2871} bulk L1$_0$
CoPt,\cite{Ravindran_prb63_144409} Co atoms on the Pt (111) surface
(experiment~\cite{Gambardella_Science300_1130} and
theory~\cite{MoscaConte_prb78_014416}), and Co dimers on the
graphite (0001) surface (our prediction) consecutively differ from
each other by about one order of magnitude.
There is no upper estimate available for Co on Pt (111), but
a value of 74 meV has been found for Co on Ag (001).\cite{Nonas01}
}
\label{fig3}
\end{figure}

\clearpage

\begin{figure}
\epsfig{file=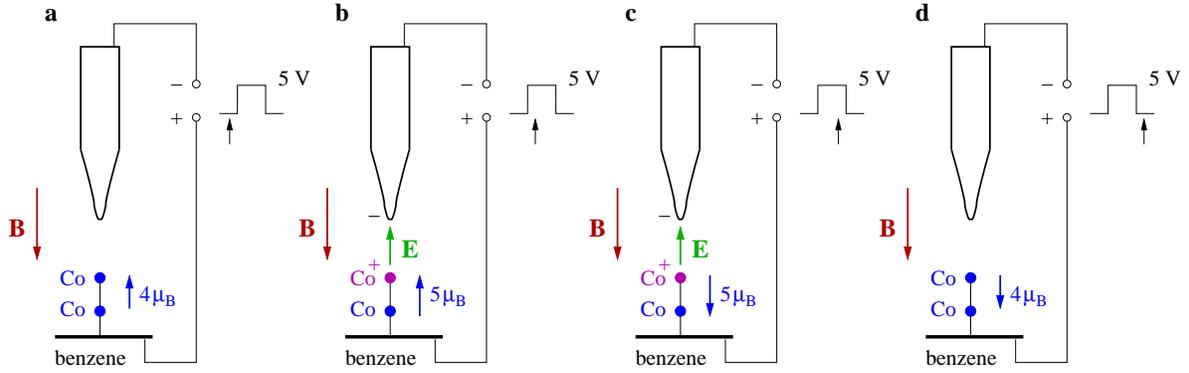, width=0.95\textwidth}
\caption{(color online)
Scheme of a bit-sensitive writing process. The time evolution from
panel {\bf a} to {\bf d} is indicated by arrows below the voltage curve.
A moderate magnetic field B is applied quasi-statically,
while a short electric field (E) pulse temporarily quenches the
anisotropy by ionizing the outer Co atom. The field is applied through
a metallic scanning tunneling microscope tip at a distance of 1 nm.
}
\label{fig4}
\end{figure}

\clearpage



\end{document}